\documentclass[aps,prl,showpacs,nofootinbib,twocolumn,
nobalancelastpage,superscriptaddress]{revtex4}
\usepackage[dvips]{epsfig}
\usepackage{graphicx}
\usepackage{amsmath}

\def\half{\frac{1}{2}}
\def\quarter{\frac{1}{4}}
\def\third{\frac{1}{3}}

\def\numu{\nu_\mu}
\def\nut{\nu_\tau}
\begin{document}


\title{Pseudo-Dirac Neutrinos, a Challenge for Neutrino Telescopes}

\author{John F. Beacom}
\affiliation{NASA/Fermilab Astrophysics Center, Fermi National Accelerator
Laboratory, Batavia, Illinois 60510-0500}

\author{Nicole F. Bell}
\affiliation{NASA/Fermilab Astrophysics Center, Fermi National
Accelerator Laboratory, Batavia, Illinois 60510-0500}
\affiliation{Kavli Institute for Theoretical Physics, University of
California, Santa Barbara 93106}

\author{Dan Hooper}
\affiliation{Department of Physics, University of Wisconsin,
Madison, Wisconsin 53706}

\author{\linebreak John G. Learned}
\affiliation{Department of Physics and Astronomy, University of
Hawaii, Honolulu, Hawaii 96822} 
\affiliation{Kavli Institute for Theoretical Physics, University of
California, Santa Barbara 93106}

\author{Sandip Pakvasa}
\affiliation{Department of Physics and Astronomy, University of Hawaii,
Honolulu, Hawaii 96822}
\affiliation{Kavli Institute for Theoretical Physics, University of
California, Santa Barbara 93106}

\author{Thomas J. Weiler}
\affiliation{Department of Physics and Astronomy, Vanderbilt University,
Nashville, Tennessee 37235}
\affiliation{Kavli Institute for Theoretical Physics, University of
California, Santa Barbara 93106}

\date{July 10, 2003}

\begin{abstract}
Neutrinos may be pseudo-Dirac states, such that each generation is
actually composed of two maximally-mixed Majorana neutrinos separated
by a tiny mass difference.  The usual active neutrino oscillation
phenomenology would be unaltered if the pseudo-Dirac splittings are
$\delta m^2 \alt 10^{-12}$ eV$^2$; in addition, neutrinoless double
beta decay would be highly suppressed.  However, it may be possible to
distinguish pseudo-Dirac from Dirac neutrinos using high-energy
astrophysical neutrinos.  By measuring flavor ratios as a function of
$L/E$, mass-squared differences down to $\delta m^2 \sim 10^{-18}$
eV$^2$ can be reached.  We comment on the possibility of probing
cosmological parameters with neutrinos.
\end{abstract}

\pacs{95.85.Ry, 96.40.Tv, 14.60.Pq \hspace{2cm} 
FERMILAB-Pub-03/201-A, MADPH-03-1337}


\maketitle


Are neutrinos Dirac or Majorana fermions?  Despite the enormous
strides made in neutrino physics over the last few years, this most
fundamental and difficult of questions remains unanswered.  The
observation of neutrinoless double beta decay would unambiguously
signal Majorana mass terms and hence lepton number violation.  If no
neutrinoless double beta decay signal is seen, it may be
tempting to conclude that neutrinos are Dirac particles, particularly
if there is independent evidence from tritium beta decay or cosmology
for significant neutrino masses.  However, Majorana mass terms 
may still exist, though their effects would
be hidden from most experiments.  Observations with neutrino
telescopes may be the only way to reveal their existence.

\begin{figure}[t]
\begin{center}
\includegraphics[width=1.8in]{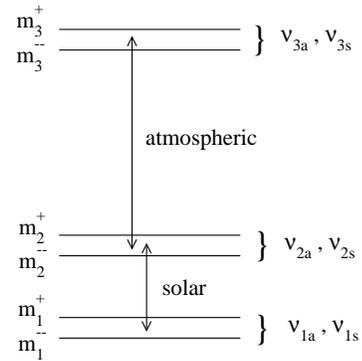}
\caption{\label{split6} The neutrino mass spectrum, showing the usual
solar and atmospheric mass differences, as well as the pseudo-Dirac
splittings in each generation (though shown as equal, we assume they
are independent).  The active and sterile components of each
pseudo-Dirac pair are $\nu_{ja}$ and $\nu_{js}$, and are maximal
mixtures of the mass eigenstates $\nu_j^+$ and $\nu_j^-$.  Neither the
ordering of the active neutrino hierarchy, nor the signs of the
pseudo-Dirac splittings, has any effect on our discussion.}
\end{center}
\vspace{-5mm}
\end{figure}

The generic mass matrix in the $\left(\nu_L, (\nu_R)^C\right)$ basis is
\begin{equation}
\left(\begin{array}{cc}
m_L & m_D    \\
m_D & m_R
\end{array}\right). 
\end{equation}
A Dirac neutrino corresponds to the case where $m_L=m_R=0$, and may be
thought of as the limit of two degenerate Majorana neutrinos with
opposite CP parity.  Alternatively, we may form a pseudo-Dirac
neutrino~\cite{pD,kobayashi} by the addition of tiny Majorana
mass terms $m_L, m_R \ll m_D$, which have the effect of splitting the
Dirac neutrino into a pair of almost degenerate Majorana neutrinos,
each with mass $\sim m_D$.  The mixing angle between the active and
sterile states is very close to maximal, $\tan(2\theta) =
2m_D/(m_R-m_L) \gg 1$, and the mass-squared difference is $\delta m^2
\simeq 2m_D(m_L+m_R)$.  For three generations, the mass spectrum is
shown in Fig.~\ref{split6}.  The mirror model can produce a very
similar mass spectrum~\cite{mirror,berezinsky}.

The current theoretical prejudice is for the right-handed Majorana
mass term to be very large, $m_R \gg m_D $, giving rise to the see-saw
mechanism.  Then the right-handed states are effectively hidden from
low energy phenomenology, since their mixing with the active states is
suppressed through tiny mixing angles.  This is desirable, since no
direct evidence for right-handed (sterile) states has been observed
(we treat both solar and atmospheric neutrinos as active-active
transitions, and do not attempt to explain the LSND~\cite{LSND} anomaly).
If right-handed neutrinos exist, where else can they hide?  An
alternative to the see-saw mechanism is pseudo-Dirac neutrinos.  Here,
although the mixing between active and sterile states is maximal, such
neutrinos will, in most cases, be indistinguishable from Dirac
neutrinos, as very few experiments can probe very tiny mass-squared
differences.

In the Standard Model, $m_D$ arises from the conventional Yukawa
couplings and hence its scale is comparable to other fermion
masses. In the see-saw model, $m_R$ is identified with some large GUT
or intermediate scale mass, and thus small neutrino masses are
achieved. For pseudo-Dirac masses, on the other hand, we need both
$m_L$ and $m_R$ to be small compared to $m_D$.  The smallness of $m_L$
with respect to $m_D$ follows from their $SU(2)_L$ properties; the
former breaks it while the latter is invariant under it. 
A similar property with respect to a $SU(2)_R$ 
(obtained with a low-energy $SU(2)_L \otimes SU(2)_R$ 
symmetry group) may also make
$m_R$ small compared to $m_D$. 
Specific examples which achieve precisely this are given in
Ref.~\cite{langacker}.  
While there still remains the problem of keeping
$m_D$ itself small enough, so that the physical neutrino masses are tiny
compared to the other fermions, there are a number of suggestions of how
this may arise~\cite{cleaver,chang,extradim}.

\begin{figure}[t]
\begin{center}
\includegraphics[width=3.25in]{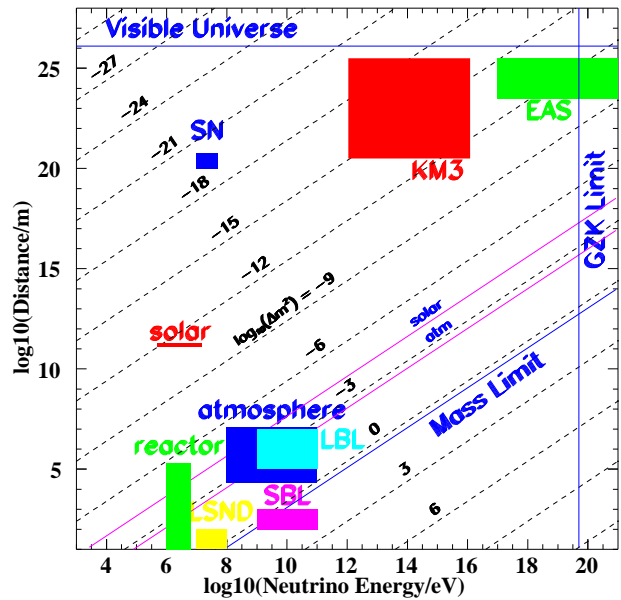}
\caption{\label{learnedplot} The ranges of distance and energy covered
in various neutrino experiments.  The diagonal lines indicate the
mass-squared differences (in eV$^2$) that can be probed with vacuum
oscillations; at a given $L/E$, larger $\delta m^2$ values can be
probed by averaged oscillations.  
The shaded regions display the sensitivity of solar, atmospheric, reactor, 
supernova (SN), short-baseline (SBL), long-baseline (LBL), LSND \cite{LSND} 
and extensive air shower (EAS) experiments.
We focus on the KM3 region, which describes the parameter space that would 
be accessible to a 1-km$^3$ scale neutrino telescope, given sufficient flux.
Current neutrino flux estimates for extragalactic sources indicate that it 
will be a challenge for km-scale experiments to make a sensitive test of the 
scenario proposed here,
and larger scale experiments would likely be necessary.
}
\end{center}
\vspace{-5mm}
\end{figure}

Astronomical-scale baselines ($L \agt E/\delta m^2$) will be required 
to uncover the oscillation effects of very tiny 
$\delta m^2$~\cite{berezinsky,GRBnus}.
Crocker, Melia, and Volkas have considered possible distortions to the
$\nu_\mu$ spectrum~\cite{roland}.
Fig.~\ref{learnedplot} shows the range of neutrino mass-squared
differences that can be probed with different classes of experiments.
Present limits on pseudo-Dirac splittings arise from the solar and
atmospheric neutrino measurements.  Splittings of less than about
$10^{-12}$ eV$^2$ (for $\nu_1$ and $\nu_2$) have no effect on the
solar neutrino flux \cite{berezinsky}, while a pseudo-Dirac splitting of 
$\nu_3$ could be as large as about $10^{-4}$ eV$^2$ before affecting the
atmospheric neutrinos.
  
Note that models with light sterile neutrinos often conflict with big
bang nucleosynthesis limits on the number of light degrees of freedom
in thermal equilibrium in the early universe.  However, the sterile 
component of each Pseudo-Dirac pair will not be populated, provided the 
mass splitting of each pair is
sufficiently small, as will be the case for the examples we consider
here.

\smallskip


{\bf Formalism.---} 
Let $(\nu_1^+, \nu_2^+, \nu_3^+; \nu_1^-,\nu_2^-,\nu_3^-)$ denote the
six mass eigenstates, where $\nu^+$ and $\nu^-$ are a
nearly-degenerate pair.  A $6 \times 6$ mixing matrix 
rotates the mass basis into the flavor basis
$(\nu_e,\numu,\nut;\nu_{e}',\nu_{\mu}',\nu_{\tau}')$.  In general, for
six Majorana neutrinos, there would be fifteen rotation angles and
fifteen phases.  However, for pseudo-Dirac neutrinos, Kobayashi and
Lim~\cite{kobayashi} have given an elegant proof that the $6 \times 6$
matrix $V_{\rm KL}$ takes the very simple form (to lowest order in
$\delta m^2/m^2$):
\begin{equation}
V_{\rm KL} = \left(\begin{array}{cc}
U & 0    \\
0 & U_R
\end{array}\right) \cdot
\left(\begin{array}{cc}
V_1 & iV_1    \\
V_2 & -iV_2
\end{array}\right),
\end{equation}
where the $3\times3$ matrix $U$ is just the usual mixing matrix
determined by the atmospheric and solar observations, the $3\times3$
matrix $U_R$ is an unknown unitary matrix, and $V_1$ and $V_2$ are the
diagonal matrices $V_1 = {\rm diag}(1,1,1)/\sqrt{2}$, and $V_2 = {\rm
diag}(e^{-i \phi_1}, e^{-i \phi_2}, e^{-i \phi_3})/\sqrt{2}$. The
$\phi_i$ are arbitrary phases.
As a result, the three active neutrino states are described in terms
of the six mass eigenstates as:
\begin{equation}
\nu_{\alpha L} = U_{\alpha j} 
\frac{1}{\sqrt{2}} \left(\nu_{j}^+ + i \nu_{j}^-  \right).
\end{equation}
The nontrivial matrices $U_R$ and $V_2$ are not accessible to active
flavor measurements.  The flavor conversion probability can thus be
expressed as
\begin{equation}
\label{prob1.5}
P_{\alpha\beta} = \frac{1}{4} \left| \sum^3_{j=1} U_{\alpha j}
\left\{ e^{i(m_j^+)^2 L/2E} + e^{i(m_j^-)^2 L/2E} \right\} 
U^*_{\beta j} \right|^2\,.
\end{equation}
The flavor-conserving probability is also given by this formula, with
$\beta = \alpha$.  Hence, in the description of the three active
neutrinos, the only new parameters beyond the usual three angles and
one phase are the three pseudo-Dirac mass differences, $\delta m^2_j
\equiv (m^+_j)^2 -(m^-_j)^2$.  In the limit that the $\delta m^2_j$
are negligible, the oscillation formulas reduce to the standard ones
and there is no way to discern the pseudo-Dirac nature of the neutrinos.

We assume that the neutrinos oscillate in vacuum.
The matter potential from relic neutrinos can affect the 
astrophysical neutrino oscillation probabilities, but only if the
neutrino-antineutrino asymmetry of the background is large, of order
1~\cite{Lunardini}.  For present limits on that asymmetry, of order
0.1~\cite{degen}, or for less extreme redshifts than assumed 
in Ref.~\cite{Lunardini}, 
matter effects are negligible.

Supernova neutrinos from distances exceeding $(E / 10 {\rm\ MeV})
(10^{-15} {\rm\ eV}^2 / \delta m^2)$ parsecs will arrive as a 50/50
mixture of active and sterile neutrinos due to vacuum oscillations.
However, we focus on the potentially cleaner signature of flavor
ratios of high-energy astrophysical neutrinos.

\smallskip


{\bf $L/E$-Dependent Flavor Ratios.---}
Given the enormous pathlength between astrophysical neutrino sources
and Earth, the phases due to the relatively large solar and
atmospheric mass-squared differences will average out 
(or equivalently, decohere).  
The neutrino density matrix $\rho$ is then
mixed with respect to the three usual mass states but coherent between
the two components of each pseudo-Dirac pair:
\begin{eqnarray}
\label{denmatrix}
\rho &=& \frac{1}{2} \sum_\alpha w_\alpha 
\sum_{j=1}^3 
|U_{\alpha j}|^2\, \Bigl\{  |\nu_j^+ \rangle\langle \nu_j^+|\,+\,
|\nu_j^- \rangle\langle \nu_j^-|
\\
&+&  i e^{-i\delta m^2_j L/2E}\, |\nu_j^- \rangle\langle \nu_j^+|
-ie^{+i\delta m^2_j L/2E}\, |\nu_j^+ \rangle\langle \nu_j^-|  \Bigr\}
\nonumber
\end{eqnarray}
Here $w_\alpha$ is the relative flux of $\nu_\alpha$ at the source,
such that $\sum_{\alpha}w_\alpha=1$.  The probability for a neutrino
telescope to measure flavor $\nu_\beta$ is then $P_\beta = \langle
\nu_\beta |\rho |\nu_\beta\rangle$, which becomes
\begin{equation}
\label{prob4}
P_\beta = \sum_\alpha w_\alpha \sum_{j=1}^3
|U_{\alpha j}|^2\, |U_{\beta j}|^2 \,
\left[1-\sin^2\left(\frac{\delta m^2_j\,L}{4E}\right)\right]\,.
\end{equation}
In the limit that $\delta m^2_j \rightarrow 0$, Eq.~(\ref{prob4})
reproduces the standard expressions.  The new oscillation terms are
negligible until $E/L$ becomes as small as the tiny pseudo-Dirac
mass-squared splittings $\delta m^2_j$.

Since $|U_{e3}|^2 \simeq 0$, the mixing matrix $U$ for three active
neutrinos is well approximated by the product of two rotations,
described by the ``solar angle'' $\theta_{\rm solar}$ and the
``atmospheric angle'' $\theta_{\rm atm} \simeq 45^\circ$.  
The pion production and decay chain at the source produces expected fluxes of
$w_e=1/3$ and $w_\mu =2/3$.  In the absence of pseudo-Dirac
splittings, it is well known~\cite{111} that this results in $P_\beta
\simeq 1/3$ for all flavors, thus the detected flavor ratios
are $\nu_e : \nu_\mu : \nu_\tau = 1:1:1$.  Here and elsewhere, this
$\nu_\mu - \nu_\tau$ symmetry is obtained when $\theta_{\rm atm} =
45^\circ$ and $U_{e3} = 0$.  If pseudo-Dirac splittings are present,
we thus expect
\begin{eqnarray}
\label{deviate1}
\delta P_\beta &\equiv& 
-\third
\left[
|U_{\beta 1}|^2\,\chi_1 +
|U_{\beta 2}|^2\,\chi_2 +
|U_{\beta 3}|^2\,\chi_3
\right]\,, 
\end{eqnarray}
where $\delta P_\beta \equiv P_\beta-\third$, and we have defined, for
shorthand,
\begin{equation}
\label{chi}
\chi_j \equiv 
\sin^2\left(\frac{\delta m^2_j\,L}{4E}\right)\,.
\end{equation}
In the absence of pseudo-Dirac terms, flavor democracy is expected.
However, the pseudo-Dirac splittings lead to an oscillatory,
flavor-dependent, reduction in flux, allowing us to test the possible
pseudo-Dirac nature of the neutrinos with neutrino telescopes.  The
signatures are flavor ratios which depend on astronomically large $L/E$.

\begin{table}[t]
\vspace{-2mm}
\caption{\label{ratios} Flavor ratios $\nu_e:\nu_\mu$ for various
scenarios.  The numbers $j$ under the arrows denote the pseudo-Dirac
splittings, $\delta m^2_j$, which become accessible as $L/E$
increases.  Oscillation averaging is assumed after each transition
$j$.  We have used $\theta_{\rm atm} = 45^\circ$, $\theta_{\rm solar}
= 30^\circ$, and $U_{e3} = 0$.}
\begin{tabular*}{0.48\textwidth}{@{\extracolsep{\fill}}ccccccc}
\hline\hline
$1:1$ & $\xrightarrow[3]{\phantom{1,2,3}}$ & $4/3:1$ &
$\xrightarrow[2,3]{\phantom{1,2,3}}$  & $14/9:1$ &
$\xrightarrow[1,2,3]{\phantom{1,2,3}}$ & $1:1$  \\
$1:1$ & $\xrightarrow[1]{\phantom{1,2,3}}$ & $2/3:1$  &
$\xrightarrow[1,2]{\phantom{1,2,3}}$  & $2/3:1$ & 
$\xrightarrow[1,2,3]{\phantom{1,2,3}}$ & $1:1$  \\
$1:1$ & $\xrightarrow[2]{\phantom{1,2,3}}$ & $14/13:1$ &
$\xrightarrow[2,3]{\phantom{1,2,3}}$  & $14/9:1$ & 
$\xrightarrow[1,2,3]{\phantom{1,2,3}} $ & $1:1$  \\
$1:1$ & $\xrightarrow[1]{\phantom{1,2,3}}$ & $2/3:1$  &
$\xrightarrow[1,3]{\phantom{1,2,3}}$  & $10/11:1$ &
$\xrightarrow[1,2,3]{\phantom{1,2,3}}$ & $1:1$  \\
$1:1$ & $\xrightarrow[3]{\phantom{1,2,3}}$ & $4/3:1$ &
$\xrightarrow[1,3]{\phantom{1,2,3}}$  & $10/11:1$ &
$\xrightarrow[1,2,3]{\phantom{1,2,3}}$ & $1:1$  \\
$1:1$ & $\xrightarrow[2]{\phantom{1,2,3}}$ & $14/13:1$ &
$\xrightarrow[1,2]{\phantom{1,2,3}}$  & $2/3:1$ &
$\xrightarrow[1,2,3]{\phantom{1,2,3}}$ & $1:1$
\vspace{1mm} \\
\hline\hline
\end{tabular*}
\vspace{-5mm}
\end{table}

As a representative value, we take $\theta_{\rm solar} = 30^{\circ}$.
Then the flavors deviate from the democratic $\third$ value by
\begin{eqnarray}
\label{deviate3}
\delta P_e &=& -\third\,\left[ \frac{3}{4}\chi_1 + \quarter\,\chi_2 \right],\nonumber\\
\delta P_\mu = \delta P_\tau &=& -\third\,\left[ \frac{1}{8}\chi_1 + \frac{3}{8}\,\chi_2 
	+\half\,\chi_3\right] \,.
\end{eqnarray}
The latter equality is due to the $\nu_\mu-\nu_\tau$ symmetry.

We show in Table~\ref{ratios} how the $\nu_e : \nu_\mu$ ratio is
altered if we cross the threshold for one, two, or all three of the
pseudo-Dirac oscillations.  The flavor ratios deviate from $1:1$ when
one or two of the pseudo-Dirac oscillation modes is accessible.  In
the ultimate limit where $L/E$ is so large that all three oscillating
factors have averaged to $\half$, the flavor ratios return to $1:1$,
with only a net suppression of the measurable flux, by a factor of
$1/2$.

It was recently pointed out that neutrino flavor ratios
will deviate significantly from 1:1:1 if one or two of the active
neutrino mass-eigenstates decay~\cite{BBHPW}.  The decay scenario
bears some resemblance to that presented here.  In
particular, if there is a range of $L/E$ values where the one or two
heavier mass states have oscillated with their pseudo-Dirac partners,
but the light state has not, then half of the heavy
states will have disappeared, to be compared with the complete
disappearance expected from unstable neutrinos~\cite{BBHPW}.  The
effects of pseudo-Dirac mass differences are much milder and
will require more accurate flavor measurements than for 
decays~\cite{BBHPW,flavor}.  
In addition, the active-active mixing angles~\cite{angles} will need to be
known independently. A detailed analysis of the prospects for measuring flavor
ratios in km-scale neutrino telescopes has been performed in Ref.\cite{flavor}.
This study shows that it will be very challenging for km-scale experiments to
sensitively test the pseudo-Dirac scenario, and larger experiments are likely 
to be necessary.

\smallskip


{\bf Neutrinoless Double Beta Decay.---}
Since the two mass eigenstates in each pseudo-Dirac pair have opposite
CP parity, no observable neutrinoless double beta decay rate is
expected.  The effective mass for neutrinoless double beta decay
experiments is given by
\begin{equation}
\langle m \rangle_{\rm eff} = 
\frac{1}{2} \sum_j U_{ej}^2 \left( m_j^+ - m_j^- \right)
= \frac{1}{2} \sum_j U_{ej}^2 \frac{\delta m^2_j}{2 m_j},
\end{equation}
which is unmeasurably small, $\langle m \rangle_{\rm eff} \alt
10^{-4}$ eV for the inverted hierarchy and even less for the normal
hierarchy.  In contrast, in the mirror model~\cite{mirror}, the sum
above has $\left( m_j^+ + m_j^- \right)$, and can thus produce an
observable signal.

\smallskip


{\bf Cosmology with Neutrinos.---} 
It is fascinating that non-averaged oscillation phases,
$\delta\phi_j=\delta m_j^2 t/4p$, and hence the factors $\chi_j$, are
rich in cosmological information~\cite{GRBnus}.  Integrating the phase
backwards in propagation time, with the momentum blue-shifted, one
obtains
\begin{eqnarray}
\delta\phi_j&=&\int_0^{z_e} dz\frac{dt}{dz}\frac{\delta m_j^2}{4p_0(1+z)}\\
	  &=&\left(\frac{\delta m_j^2 H^{-1}_0}{4p_0}\right)\,
		\int_1^{1+{z_e}}\frac{d\omega}{\omega^2}
		\frac{1}{\sqrt{\omega^3\,\Omega_m+(1-\Omega_m)}}\,,\nonumber
\label{phase}
\end{eqnarray}
where $z_e$ is the red-shift of the emitting source, and $H_0^{-1}$ is
the Hubble time, known to 10\%~\cite{keyproject}.  This result holds
for a flat universe, where $\Omega_m+\Omega_\Lambda=1$,
with $\Omega_m$ and $\Omega_\Lambda$ the matter and vacuum energy
densities in units of the critical density.
The integral
is the fraction of the Hubble time available for neutrino transit.
For the presently preferred values $\Omega_m=0.3$ and
$\Omega_\Lambda=0.7$, the asymptotic ($z_e\rightarrow\infty$) value of
the integral is 0.53.  This limit is approached rapidly: at
$z_e=1\,(2)$ the integral is 
77\% (91\%) saturated.  
For cosmologically distant ($z_e \agt 1$) sources such as gamma-ray bursts, 
non-averaged oscillation data would, in principle, allow one to
deduce $\delta m^2$ to about 20\%, without even knowing the source
red-shifts.  
Known values of $\Omega_m$ and $\Omega_\Lambda$ might 
allow one to infer the source redshifts $z_e$, or vice-versa.

Such a scenario would be the first measurement of a cosmological
parameter with particles other than photons.  An advantage of
measuring cosmological parameters with neutrinos is the fact that
flavor mixing is a microscopic phenomena and hence presumably free of
ambiguities such as source evolution or standard candle
assumptions~\cite{GRBnus,stodolsky}.  Another method of
measuring cosmological parameters with neutrinos is given in
Ref.~\cite{choubey}.


\smallskip

{\bf Conclusions.--- }
Neutrino telescope measurements of neutrino flavor ratios may achieve 
a sensitivity to mass-squared differences as small as $10^{-18}$ eV$^2$.
This can be used to probe possible tiny 
pseduo-Dirac splittings of 
each generation, and thus reveal Majorana mass terms (and lepton number 
violation) not discernable via any other means.

\smallskip

{\it Note added:} As this work was being finalized, a paper 
appeared which addresses some of the
issues herein~\cite{keranen}.


\smallskip

{\bf Acknowledgments.---} 
We thank Kev Abazajian for discussions.  J.F.B. and N.F.B. were supported by
Fermilab (under DOE contract DE-AC02-76CH03000) and by NASA grant NAG5-10842.
D.H was supported by DOE grant DE-FG02-95ER40896, J.G.L. and S.P. by DOE grant
DE-FG03-94ER40833 and T.J.W. by DOE grant DE-FG05-85ER40226.


\end{document}